# Wedging Transfer of Nanostructures

Grégory F. Schneider, Victor E. Calado, Henny Zandbergen, Lieven M.K. Vandersypen and Cees Dekker*

*Kavli Institute of Nanoscience, Lorentzweg 1, 2628 CJ Delft, The Netherlands*

*Corresponding author: c.dekker@tudelft.nl

**ABSTRACT**  We report a versatile water-based method for transferring nanostructures onto surfaces of various shapes and compositions. The transfer occurs through the intercalation of a layer of water between a hydrophilic substrate and a hydrophobic nanostructure (for example, graphene flakes, carbon nanotubes, metallic nanostructures, quantum dots, etc) locked within a hydrophobic polymer thin film. As a result, the film entrapping the nanostructure is lifted off and floats at the air-water interface. The nanostructure can subsequently be deposited onto a target substrate by the removal of the water and the dissolution of the polymeric film. We show examples where graphene flakes and patterned metallic nanostructures are precisely transferred onto a specific location on a variety of patterned substrates, even on top of curved objects such as microspheres. The method is simple to use, fast, and does not require advanced equipment.

**TOC**

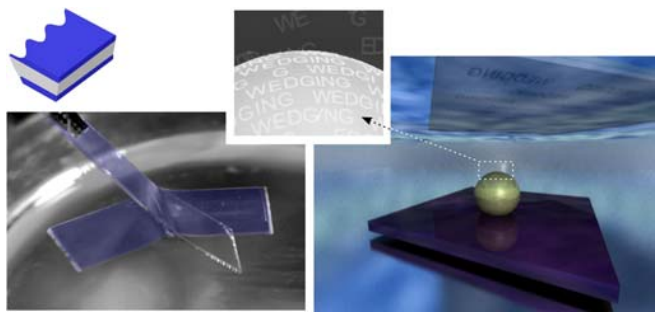

**MANUSCRIPT**

In applications ranging from soft lithography[1] to three-dimensional electronics,[2] the controlled transfer of nanostructures from one substrate to another is crucial.[3] Despite the fact that nanofabrication methods have advanced substantially,[4] there is no general and robust approach available for transferring and precisely aligning (nano)structures onto a specific device with submicrometer precision. Existing transfer methods are often limited in scope because they suffer from process-specific drawbacks, such as optimizing the pressing force *versus* chemical adhesion (for transfer printing by stamping),[5-8] the use of reactive chemicals (in lift-off by etching),[9-13] the use of irreversible mechanical stripping methodologies (e.g. peeling)[14-16] to separate two solid surfaces brought in contact, or exposure to high temperatures needed for baking polymer sticking layers or for releasing thermal adhesives.[17] This list is not exhaustive but is representative of the most popular technique used for transferring.

The new technique that we introduce here, wedging transfer, is versatile, suitable for transfer onto many different type of substrates including curved surfaces, and intrinsically combines transfer and alignment. The technique is based on the simplest view of the hydrophobic effect:[18,19] water wets hydrophilic surfaces and avoids hydrophobic ones,[20-22] and is inspired from methods generally used to prepare polymer-coated transmission electron microscopy grids.[23] As we will demonstrate, this implies that water can lift off a hydrophobic object from a hydrophilic substrate. For instance, a graphene monolayer flake can be lifted off and transferred from one substrate to another one, simply by using water as the transfer-active component.

Figure 1 depicts the process of wedging transfer. A microscope glass slide is coated on both sides with a cellulose acetate butyrate polymer thin film (see methods). Dipping the slide into water releases the polymer film from the glass slide. As a result, the polymer film floats on water. The wedging is observed to work for at least two incidence angles, here ~30° and ~150° between the polymeric films and the water meniscus. The wedging process is found to be intrinsically reversible: by retracting the glass slide again from water, the polymer film is re-deposited on the initial substrate (data not shown).

The physical driving force yielding the lift-off is the capillary force exerted by water that invades the hydrophilic/hydrophobic interface, resulting in the dynamic separation of both surfaces. The wedging only depends on the kinetics and thermodynamics of water invasion, and is therefore independent of the experimentalist skills: wedging will always occur if the experiment is performed slowly enough (few seconds as a time scale).

The wedging principle can be used for transferring nanostructures, as schematically illustrated in Figure 2. A hydrophobic pattern (e.g., the word "WEDGING") that has been fabricated on a hydrophilic substrate (such as glass, quartz, mica, $SiO_2$) is dipped into a solution of a hydrophobic polymer (cellulose acetate butyrate, 30 mg/mL in ethyl acetate, Sigma-Aldrich), thus forming a solid polymer layer covering the pattern after the evaporation of the solvent at room temperature. Subsequently, the polymer film which now includes the pattern that is to be transferred is wedged in water and floats at the water/air interface because of surface tension forces. Transfer to the new location occurs when the water level is lowered enough for the polymer film to reach the receiver substrate. Once deposited onto a selected target spot of the receiver substrate (target markers shown in black in Figure 2), the polymer film is removed by dissolving it into the solvent used for its initial dissolution (here ethyl acetate).

As a first demonstration of our method, we transferred a flake of graphene, a material which has recently received a lot of attention.[24-30] While it is now becoming routine to obtain single-layer graphene by mechanical exfoliation of bulk graphite and to identify the single-layer flakes by means of their optical contrast against the bare substrate,[31] one of the major drawbacks of this technique is that graphene flakes are arbitrarily distributed on the substrate. It would be of great use if specific flakes could be picked up from the substrate, and redeposited in a precise location on another substrate, for instance on top of prefabricated contacts, a TEM grid, or other fabricated nanostructures. Graphene thus is an excellent model material to demonstrate the applicability of the wedging transfer technique.

The wedging transfer and alignment of a graphene sample onto a target spot on the receiver substrate is depicted in Figure 3. In our setup (Figure 3A), we laterally moved the graphene flake (Figure 3B) along the water/air interface by using a probe needle that contacts the polymer layer and thereby aligned the flake over the target spot (square pattern with a round hole of 5 µm in diameter, Figure 3C). The probe needle is manually or electrically moved in the lateral and height directions with three orthogonal micrometric screws. Because of the high transparency of the film, we can use an optical microscope to align the position of the graphene flake with respect to the target with sub-micrometer accuracy. During the aligning, the water level in the Petri-dish is lowered using a syringe pump. Upon pumping out the water, the graphene is deposited onto the target spot of the receiver substrate. Its shape is preserved during the transfer (Figure 3D), and, importantly, also after the dissolution of the polymeric scaffold (Figure 3E).

For example, we performed so far 49 wedging transfers with graphene. All flakes were always successfully transferred onto SiN and $SiO_2$ receiver substrates (Figure 4A), and also onto mica surfaces. Mica is difficult to use as a substrate for graphene with the mechanical exfoliation method,[32] since layers of mica will also exfoliate when using the scotch tape. Figure 4B shows the successful wedging transfer of graphene onto mica, observed in the transmission mode of an optical microscope.[33]

We have explored the applicability of the wedging transfer to a variety of other structures. Figures 4C and 4D show examples of gold microelectrodes and gold nanostructures respectively. We observed that the hydrophobic polymer does not stick well enough to the gold material, thus resulting in transfer yields lower than 10% (data not shown). If the polymer solution, however, is supplemented with an aliphatic thiol (0.1 %v of 1-dodecanethiol, Fluka Analytical Ref. #44130), a hydrophobic self-assembled monolayer will spontaneously form onto the gold surface[34]. This hydrophobic self-assembled monolayer increases the sticking between the polymer and the gold by means of the hydrophobic interactions. This enabled us to transfer again with a success rate of 100%, gold microelectrodes patterned on $Si/SiO_2$ wafers (without chromium or titanium sticking layers), onto substrates such as $SiO_2$ already covered with other microelectrodes (Figure 4C). We also successfully transferred gold nanofabricated letters (70 nm line width, also in the presence of thiolates), onto highly curved objects such as polystyrene microspheres (Figure 4D). Some nanostructures (~10%) were not transferred fully (note some missing letters), possibly because the polymer film is not elastic enough to conform totally to the curved object. Increasing the elasticity of the film can possibly be achieved by lowering the concentration of the polymer in the solvent or adding plasticizers molecules to the polymer.

Wedging transfer has the potential to become a widely used method in nanotechnology, as it offers important advantages over the state of the art. The high success rates, versatility, tunability, and reversibility of the method can be easily checked by direct visual inspection. As a final illustration of the extreme simplicity of the method, one can take a white board marker, write a line onto a microscope slide or a stainless-steel spoon, dip it into a glass of water and immediately see the line wedging and floating on top of water. Subsequently, the line can be redeposited with a backwards movement onto the original substrate, or onto another material. From our knowledge, there are no other transfer techniques that are so effective and straightforward to use.

**Methods**

**Preparation for wedging transfer.** A hydrophilic wafer is dipped in a solution of cellulose acetate butyrate (~30 mg/mL in ethyl acetate). After dipping (the total dipping procedure lasts about three seconds), the substrate is removed from the polymer solution and the solvent is left to evaporate under ambient conditions. At this point in time, the edges of the substrate are still covered with the polymer (hindering the water to intercalate at the hydrophilic/hydrophobic interface).[23] Therefore, by using a cotton swab impregnated with the same solvent as the one used to solubilise the polymer, the edges are cleaned so that water can intercalate at both interfaces. Alternatively one can use a sharp razor blade and scratch the polymer at the edges of the wafer. Both methods yield indistinguishable results. The substrate is then dipped into water with an incidence angle of about 30°.

**Preparation of graphene samples.** We prepared graphene sheets on clean and freshly plasma oxidized Si/SiO$_2$ substrates (O$_2$, Diener) by mechanical exfoliation of natural graphite (NGS graphit) with blue NITTO tape (SPV 224P). To render graphene monolayers visible, we used Si/SiO$_2$ wafers with a 285 nm thermally grown SiO$_2$ layer (Nova Electronic Materials LTD). We located the single and few layer graphene sheets under an optical microscope and identified the number of layers by their optical contrast.[31] To ensure the intercalation of water between the substrate and the polymer, especially if the graphene flakes were prepared more than a few hours before being wedged, we usually remove junk graphite with a cotton swab impregnated with ethanol from the substrate. Only the region of interest is left untouched and is further covered with a small drop of the polymer (~2 μL). This now acts as a protection mask once the wafer is exposed to air plasma (5s, SPI Plasma Prep II). The protective mask is then dissolved in the polymer solution (30 dips are enough to solubilize the mask) and the procedure above (e.g., preparation for the wedging transfer) is performed.

**Nanofabrication of the gold 'WEDGING' patterns and microelectrodes.** A 60x60 μm$^2$ field with many copies of the word 'WEDGING' (letter line width is 70 nm) is written by electron beam lithography in a 495K/950K double spun PMMA layer. A 20 nm gold layer is evaporated. Subsequently, the PMMA/Au layer is gently lifted-off in acetone. Prior to dipping the wafer in the cellulose acetate butyrate solution (supplemented with 0.1 v% of 1-dodecanethiol), the wafer was plasma oxidized under an $O_2$ plasma for thirty seconds (SPI plasma cleaner). The patterns are then wedged onto 10 µm polystyrene microspheres (Polybead Polystyrene, Cat# 17136, lot #597574, Polysciences, Inc.) previously adsorbed from solution on a $SiO_2$ substrate. The SEM image is taken with a FEI XL30S microscope operating at 30 kV. The same method was used to fabricate the gold microelectrodes.

**Acknowledgments.**

This research was funded by the Netherlands Organisation for Scientific Research (NWO) and the Foundation for Fundamental Research on Matter (FOM).

**Figures Captions and Figures.**

**Figure 1.** Time-lapse optical imaging of the simultaneous wedging of two hydrophobic polymer layers that were deposited on both sides of a microscope glass cover slip.

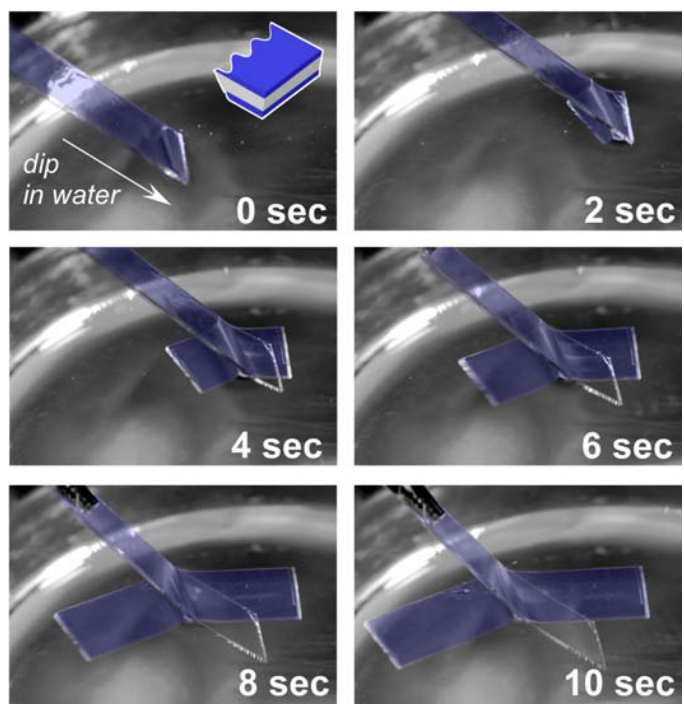

FIGURE 1

**Figure 2.** Illustration of the handling steps involved in the wedging transfer technique.

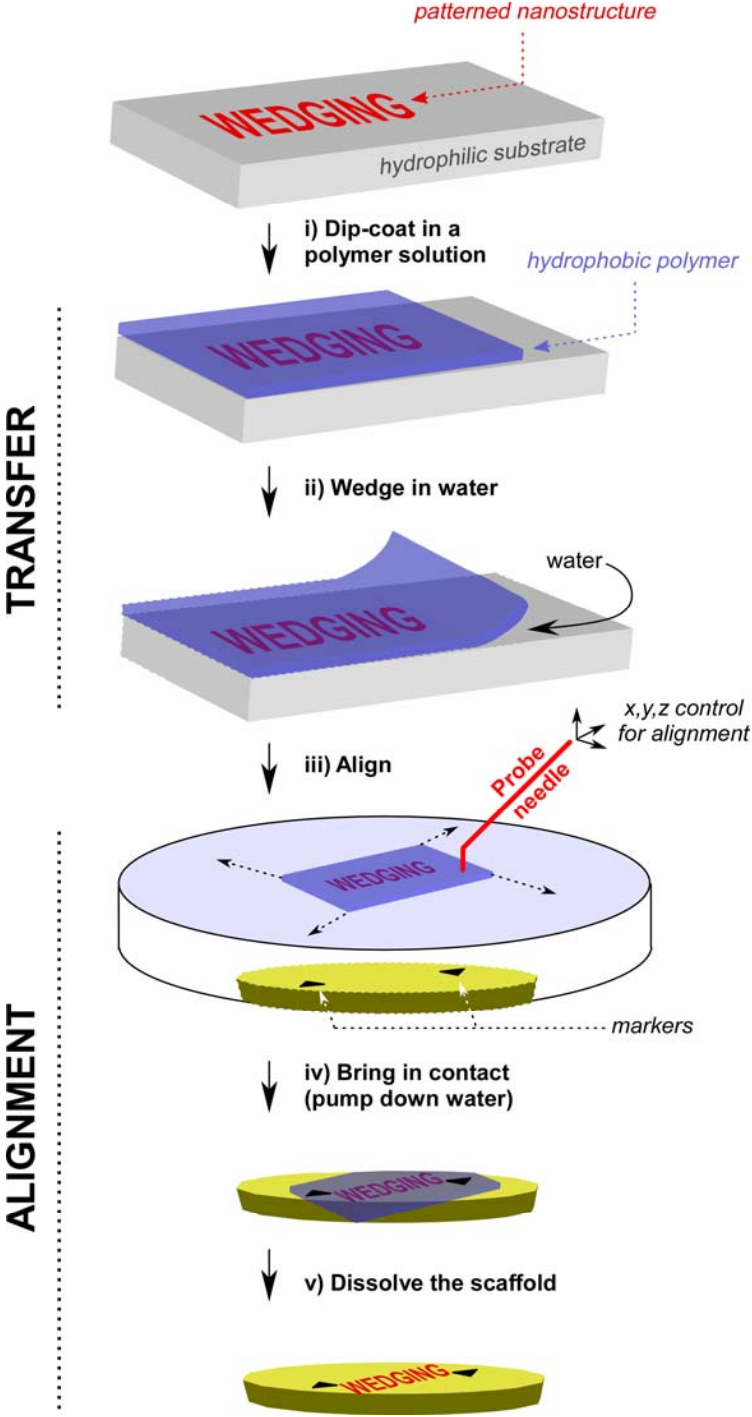

FIGURE 2

**Figure 3.** Demonstration of the wedging transfer technique with the controlled transfer of a multilayer graphene flake. A) The experimental setup consists of a conventional low-magnification optical microscope, and a standard sewing needle connected to a micrometric screw to precisely align the wedged polymer film (including the graphene pattern) on top of a target device. B) Optical microscope image of a few graphene layers on a Si/SiO$_2$ wafer obtained through the standard mechanical-exfoliation method (oxide thickness of 285 nm). C) The target device which is a SiN membrane with a 5 µm diameter hole. D) Graphene/polymer layer deposited onto the target device after the wedging and drying under ambient conditions. E) Graphene flake on the target device after dissolution of the hydrophobic polymer film. Scale bars are 10 µm in all images.

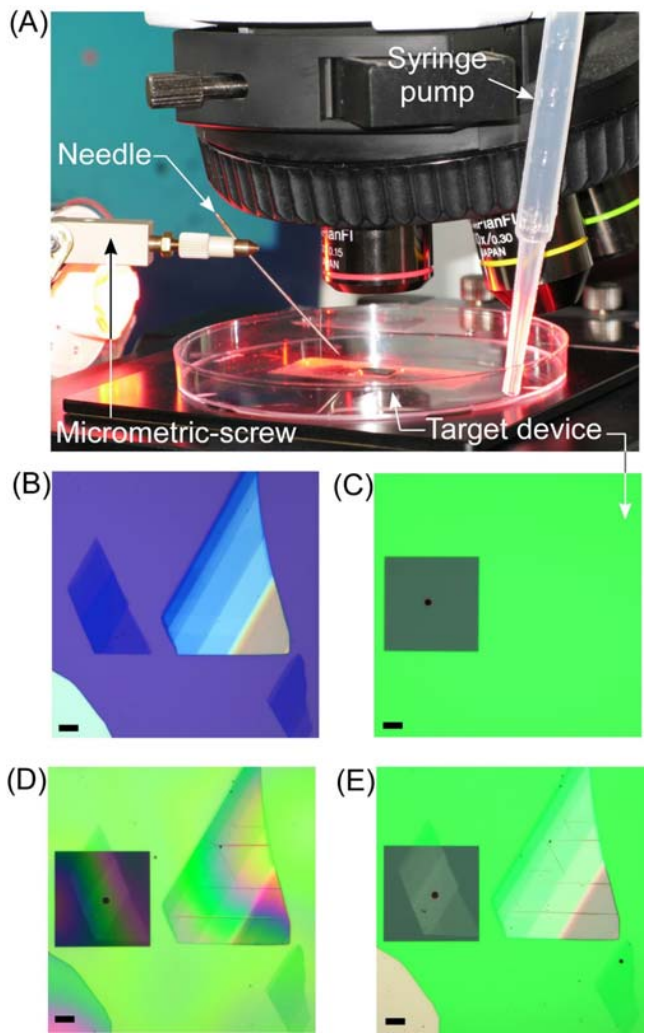

FIGURE 3

**Figure 4.** Four examples of wedging transfer of nanostructures. A) Wedging transfer of a graphene monolayer from a Si/SiO$_2$ wafer (left) onto a SiN membrane with a patterned 5 µm hole. The polymer film has not yet been dissolved (right micrograph) to enhance the contrast and the optical visibility of the monolayer onto the target substrate. Scale bar represents 10 µm. B) Wedging transfer of graphene multilayers (including a monolayer, see dashed lines pointed near the arrow) from a Si/SiO$_2$ wafer (left) onto mica, observed in the transmission mode of the optical microscope (right). Scale bar represents 10 µm. C) Scanning electron microscope image of patterned gold microelectrodes (left, smallest lines 2 µm wide, 50 nm thick, without titanium sticking layer). These electrodes were transferred from a Si/SiO$_2$ wafer onto another Si/SiO$_2$ wafer that contained a previous pattern of perpendicularly oriented microelectrodes (smallest line 200 nm wide, 20 nm Ti, 50 nm gold). Scale bar represents 10 µm. D) Scanning electron microscope image of the letters 'WEDGING' that have been transferred from Si/SiO$_2$ onto a 10 µm diameter polystyrene microsphere (the scale bar is 5 µm). An enlargement of the square is shown on the right (the scale bar is 1 µm). The gold letters were made by e-beam lithography (see methods) and have 70 nm line width and 20 nm thickness.

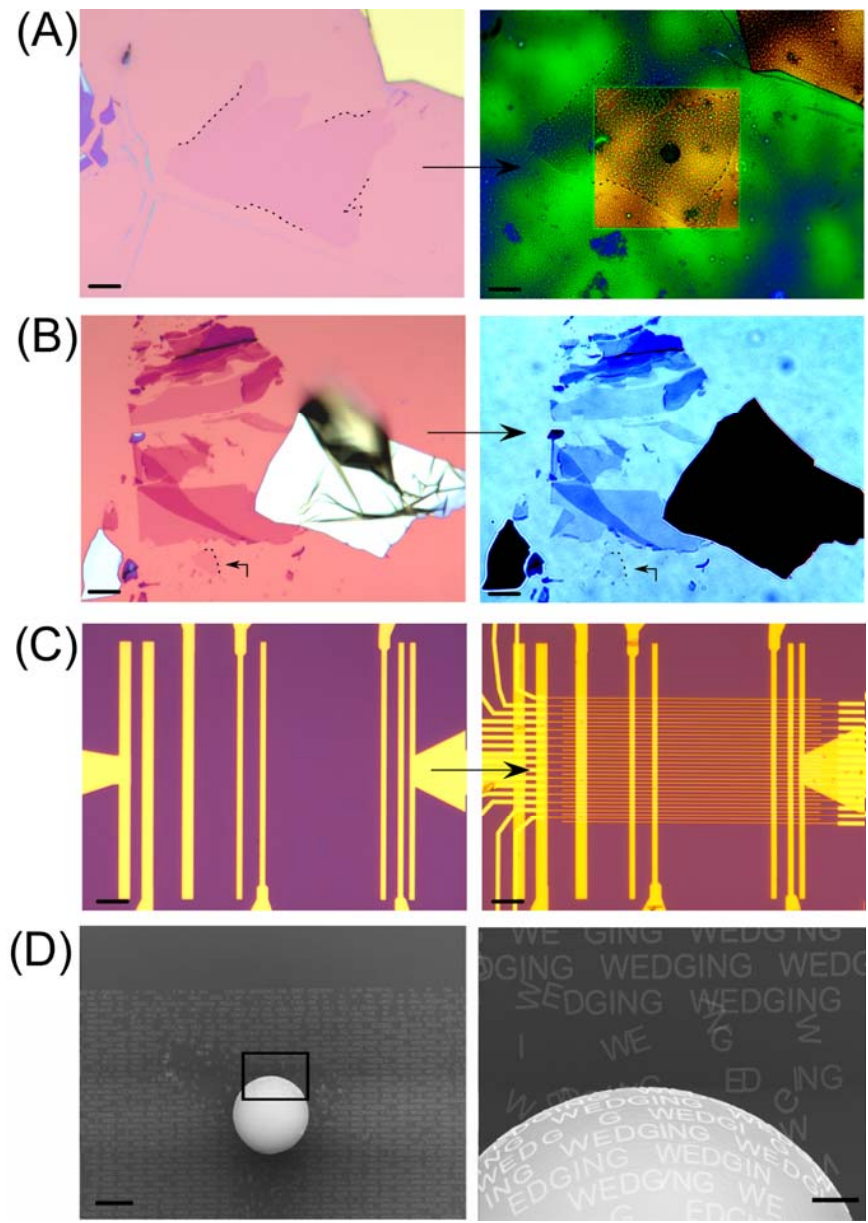

FIGURE 4